\documentclass[a4paper,11pt]{article}

\usepackage{jcappub}

\usepackage{slashed}
\usepackage{youngtab}

\newcommand\fverb{\setbox\fverbbox=\hbox\bgroup\verb}
\newcommand\fverbdo{\egroup\medskip\noindent%
            \fbox{\unhbox\fverbbox}\ }
\newcommand\fverbit{\egroup\item[\fbox{\unhbox\fverbbox}]}
\newbox\fverbbox
\newcommand{\nablaslash}{\not{\hbox{\kern-3pt $\nabla$}}}
\newcommand{\nn}{\nonumber}

\title{Massive gravitational waves in Chern-Simons modified gravity}

\author{Yun~Soo~Myung}
\author{and~Taeyoon~Moon}
\affiliation{Institute of Basic Science and Department of Computer
Simulation, Inje University,\\
Gimhae 621-749, Korea}

\emailAdd{ysmyung@inje.ac.kr} \emailAdd{tymoon@inje.ac.kr}

\abstract{We consider the nondynamical Chern-Simons (nCS) modified
gravity, which is regarded as a parity-odd theory of massive gravity
in four dimensions. We first find polarization modes of
gravitational waves for $\theta=x/\mu$ in nCS modified gravity by
using the Newman-Penrose formalism where the null complex tetrad is
necessary to specify gravitational waves. We show that in the
Newman-Penrose formalism, the number of polarization modes is one in
addition to an unspecified $\Psi_4$, implying three degrees of
freedom for $\theta=x/\mu$.  This compares with two for a canonical
embedding of $\theta=t/\mu$. Also, if one introduces the Ricci
tensor formalism to describe a massive graviton arising from the nCS
modified gravity, one finds one massive mode after making
second-order wave equations, which  is compared to five found from
the parity-even Einstein-Weyl gravity.}

\begin{document}

\maketitle \flushbottom
\section{Introduction}

Topologically massive gravity (TMG) including a gravitational
Chern--Simons term (gCS) is a three-dimensional gravity
theory~\cite{DJT, DJT2} with a massive propagating degree of freedom
(DOF). Since the gCS term is odd under parity, the theory shows a
single DOF  of a given helicity, whereas the other helicity mode
remains massless. Li, Song, and Strominger~\cite{LSS} have shown
that the third-order Einstein equation (the first-order equation for
the linearized Einstein tensor $\delta G_{\mu\nu}$) of cosmological
TMG is changed into the same first-order equation for a massive
graviton $h_{\mu\nu}$ when one chooses the transverse-traceless
gauge. This shows clearly that the cosmological TMG is regarded as
the first-order gravity theory. Later on, Bergshoeff, Hohm, and
Townsend have proposed new massive gravity (NMG)  by adding a
quadratic curvature term to the Einstein-Hilbert action~\cite{bht}.
Although this term was designed to reproduce  the Fierz-Pauli action
for a massive graviton, there is no way to avoid ghost states in
higher dimensions ($D\geq4$), because of the fourth-order theory in
the metric tensor formalism~\cite{Bergshoeff:2013vra}. Since the NMG
preserves parity unlike the TMG, a massive graviton acquires the
same mass for both helicity states, showing two DOF.  However, we
did not know what are polarization modes of gravitational waves
(GWs) in TMG and NMG. Since two theories belong to higher-order
gravity in three dimensions, we need to introduce the Newman-Penrose
(NP) formalism~\cite{Newman} where the null real tetrad is necessary
to specify a few polarization modes of GWs.   It turned out that one
mode is $\Phi_{12}$ for the TMG, whereas two modes are $\Phi_{22}$
and $\Phi_{12}$ for the NMG~\cite{Moon:2011gg}. This shows a perfect
matching between the Eistein tensor~\cite{bht} and NP formalisms
because in three dimensions, only the massive graviton exists and
the Weyl tensor vanishes.

On the other hand, there is no consensus on the number of graviton
DOF in nondynamical Chern-Simons
(nCS)~\cite{Jackiw:2003pm,Grumiller:2007rv,Moon:2011ef} and
dynamical Chern-Simons (dCS) modified
gravity~\cite{Yunes:2009hc,Alexander:2009tp,Cardoso:2009pk,Moon:2011fw}
in four dimensions, because it depends on the (non)dynamical CS as
well as the choice of $\theta$. The number of graviton DOF
propagating around a Minkowski background was two ($h_{ij}^{\rm
TT}$) as general relativity for the canonical CS coupling of
$v_\mu=\partial_\mu\theta=(1/\mu,\bf{0})$~\cite{Jackiw:2003pm}.
Choosing a class of exact solutions describing plane-fronted
gravitational waves ($pp$-waves) along $+z$ axis, the graviton is
also  described by two DOF
($\Psi_4=-R_{n\tilde{m}n\tilde{m}}=\ddot{h}_{+}-i\ddot{h}_{\times}$)
for the nCS gravity~\cite{Grumiller:2007rv}, while it has three
($\Psi_4$ and
$\Phi_{22}=-R_{nmn\tilde{m}}=\frac{1}{2}\ddot{h}_{m\tilde{m}}$) for
the dCS gravity~\cite{Sopuerta:2009iy}. Here $h_{+,\times}$ are the
plus/crossing polarizations of the waveform and the overhead dot
($\dot{}$) denotes the differentiation with respect to the retarded
null coordinate $u=t-z$. However, in the weak amplitude regime of
$\theta^2 \simeq 0$, the effect of $\Phi_{22}$ is negligible, which
establishes $\Psi_4$ as in general relativity.

At this stage, we should mention the difference between the
canonical embedding $\theta=t/\mu$ in Minkowski
background~\cite{Jackiw:2003pm} and a spacelike choice
$\theta=x/\mu$ in Minkowski background~\cite{Boldo:2009ui} and
AdS$_4$ background~\cite{Moon:2011ef}.  As was emphasized
previously, there is no additional degrees of freedom for the
canonical embedding because this does not induce time derivative but
space derivative additionally. However, a spacelike choice of either
$\theta=x/\mu$ or $\theta=y/\mu$ provides a time derivative in
addition to second-order time derivative. The latter mimics the TMG
in three dimensions.  Recently, it was shown that six and ten
initial conditions are needed to specify the time evolution of
physical perturbations for  $\theta=\mu t+\Psi(r)$ nCS and dCS
theories  in the Schwarzschild black hole
spacetime~\cite{Motohashi:2011ds}. Explicitly, noting that the
Einstein-Hilbert term provides four, the dynamical CS scalar
$\theta$ has two, and the CS term brings four, we have ten in the
dCS gravity.  However, one has six for the nCS gravity because the
Pontryagin constraint ${}^{*}RR=0$ kills two in eight. Actually,
this corresponds to three and five DOF for the nCS and dCS gravity,
respectively. This supports that the number of DOF is three for the
nCS gravity when one uses the metric tensor formalism as well as
$\theta=\mu t+\Psi(r)$ with $\Psi(r)\not=0$. Actually, the number of
propagating DOF is independent of a given background.

Even though a quadratic gravity required null complex tetrad to
specify six independent polarization modes of $\{ \Psi_2, \Psi_3,
\Psi_4, \Phi_{22}\}$ in the NP
formalism~\cite{Eardley:1974nw,Alves:2009eg}, the Einstein-Weyl (EW)
gravity
\begin{equation}\label{ewaction}
S_{\rm EW}=\frac{1}{2\kappa^2}\int
d^4x\sqrt{-g}\Big[R+\frac{1}{\mu^2}\Big(R_{\mu\nu}R^{\mu\nu}-\frac{1}{3}R^2\Big)\Big]
\end{equation}
has seven modes $\{ \Psi_3, \Psi_4, \tilde{\Psi}_4, \Phi_{22}\}$
with $\tilde{\Psi}_4$ the massive ghost modes because the EW gravity
is a fourth-order theory in the metric
formalism~\cite{Myung:2014jha}. Here $\Psi_3$, $\Psi_4$, and
$\tilde{\Psi}_4$ are complex, providing six DOF. Analyzing the
rotational behavior of the former set shows the respective helicity
values $s=\{0,\pm1,\pm2,0\}$. However, $\Psi_4$ always remains
unspecified here because it must be determined by the Riemann
tensor. This implies that the NP formalism is not appropriate for
describing both massless and massive modes with seven DOF totally.
To account for a massive mode only, it would be better to introduce
the Ricci tensor formalism without ghost problem, instead of the
metric tensor formalism with ghost problem.

Accordingly, it is urgent to clarify how many DOF  are  in nCS
modified gravity (\ref{csaction}) when the nCS modified gravity with
$\theta=x/\mu$ is considered as a massive gravity.  In order to find
it, we first propose the similarity and difference between 3D and 4D
massive gravities based on the metric tensor formalism:

\vspace{0.5cm}
\begin{tabular}{|c|c|c|c|}
  \hline
  dimensions & parity-odd  gravity  & parity-even gravity  & remarks \\
  \hline
  3D & TMG (1DOF)& NMG (2) & massive mode\\ \hline
  4D & nCS gravity (3) & EW gravity (7)& massless and massive modes\\
  \hline
\end{tabular}
\vspace{0.5cm}

 We emphasize that the $\theta=x/\mu$ nCS modified gravity was
considered as a 4D extension of the TMG, whereas the dCS modified
gravity has been formulated by treating the CS scalar $\theta$ as a
dynamical field. Hence, in this work, we investigate polarization
modes of gravitational waves in the nCS modified gravity by using
the NP formalism where the null complex tetrad is necessary to
specify gravitational waves. The linearized Einstein equation
corresponds to the first-order equation for the linearized Ricci
tensor which will be used to kill the independent polarization modes
propagating on the Minkowski spacetime.  It turns out that the
number of polarization modes is three  for the nCS case by taking
into account an unspecified $\Psi_4$ which corresponds to three in
the metric tensor formalism.  If one uses the Ricci tensor formalism
in the nCS modified gravity, there is no ghost problem because it
lowers the third-order equation to the first-order one. Using the
linearized TMG and NMG~\cite{LSS,bht}, we could develop the
linearized Einstein equation which is a second-order wave equation.
This might describe a tensor wave propagating along $+z$ axis. In
this case, we obtain one DOF for a massive graviton propagating on
the Minkowski background.

\section{The Newman-Penrose formalism}

Let us first consider GWs propagating in the $+z$ direction for
simplicity. In this case, all waves are functions of $t$ and $z$. At
any point $P$, the null complex tetrad vectors $\{ {\bf
k},\bf{n},\bf{m},\bf{\tilde{m}}\}$   are related to the Cartesian
tetrad vectors $\{{\bf e}_{t},{\bf e}_{x},{\bf e}_{y},{\bf e}_{z}\}$
in  four dimensions with metric signature $(-,+,+,+)$ as
\begin{equation} \label{null-t}
{\bf k}/{\bf n}=\frac{1}{\sqrt{2}}({\bf e}_t\pm{\bf e}_z),~~ {\bf
m}/{\bf \tilde{m}}=\frac{1}{\sqrt{2}}({\bf e}_x\pm i{\bf e}_y),
\end{equation}
where they  satisfy the relations
\begin{eqnarray}
-{\bf k}\cdot{\bf n}={\bf m}\cdot {\bf \tilde{m}}=1,~~~ {\bf
k}\cdot\{{\bf m},{\bf \tilde{m}}\}={\bf n}\cdot\{{\bf m},{\bf
\tilde{m}}\}={\bf k}^2={\bf n}^2=0.
\end{eqnarray}
We note that a tensor ${\bf T}$  can be written as
\begin{eqnarray}\label{tensor}
T_{abc...}=T_{\mu\nu\rho...}a^{\mu}b^{\nu}c^{\rho_{...}},
\end{eqnarray}
where Latin indices $(a,b,c,...)$ run over  $( {\bf
k},\bf{n},\bf{m},\bf{\tilde{m}})$, while Greek indices
$(\mu,\nu,\rho,...)$ run over $(t,x,y,z)$ because we are working
with Cartesian coordinates to specify the Minkowski background.
 It is well-known that the
Weyl tensor has ten components  in four dimensions. Therefore, the
Riemann tensor with twenty  components can be decomposed into the
Ricci tensor with ten and Weyl tensor with ten as
\begin{eqnarray}
R_{\rho\sigma\mu\nu}=\Big(g_{\rho[\mu}R_{\nu]\sigma}
-g_{\sigma[\mu}R_{\nu]\rho}\Big)-\frac{1}{3}Rg_{\rho[\mu}g_{\nu]\sigma}+C_{\rho\sigma\mu\nu}.
\end{eqnarray}
Then, the NP quantities of five complex Weyl scalars $\Psi$'s
[$C_{\rho\sigma\mu\nu}$], nine $\Phi$'s[$R_{\mu\nu}$] and
$\Lambda$[$R$] represent irreducible parts of the Riemann tensor
$R_{\rho\sigma\mu\nu}$.

\begin{figure*}[t!]
   \centering
   \includegraphics[width=.63\linewidth,origin=tl]{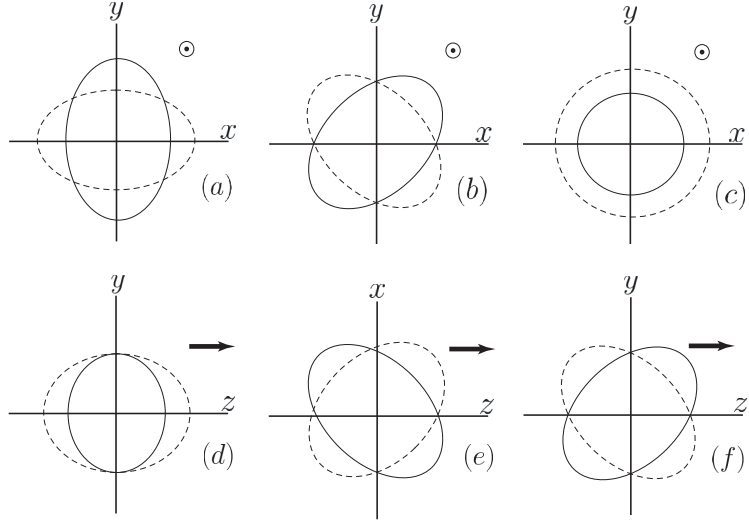}
\caption{Six polarization modes of weak, plane GW permitted in
four-dimensional gravity, which indicate $(a):~{\rm
Re}[\Psi_4],~(b):~{\rm
Im}[\Psi_4],~(c):~\Phi_{22},~(d):~\Psi_2,~(e):~{\rm
Re}[\Psi_3],~(f):~{\rm Im}[\Psi_3]$. All modes $(a)\sim(f)$ are
propagating in the $+z$ direction. The solid displacement shows that
each mode induces on a sphere of test particles, while the dashed
displacement indicates that each mode induces on a sphere of test
particles after half-period.} \label{Rnn}
\end{figure*}

 Choosing nearly plane waves propagating $+z$
direction reduces $R_{\rho\sigma\mu\nu}$ to six nonvanishing
components representing by a set of $\{ \Psi_2, \Psi_3, \Psi_4,
\Phi_{22}\}$ in the generic metric theory~\cite{Eardley:1974nw}. The
first and last ones are real scalars, while the second and third are
complex scalars. Fig. 1 shows the six polarization modes of weak,
plane, null GW permitted in any metric theory of gravity. Using the
tetrad basis, the NP quantities are represented by the Riemann
tensor and Ricci tensor as
\begin{eqnarray}
\Psi_{2}&=&-\frac{1}{6}R_{nknk}=-\frac{1}{6}R_{nk},\nn\\
\Psi_{3}&=&-\frac{1}{2}R_{nkn\tilde{m}}=-\frac{1}{2}R_{n\tilde{m}},\nn\\
\Psi_{4}&=&-R_{n\tilde{m}n\tilde{m}},\nn\\
\Phi_{22}&=&-R_{nmn\tilde{m}}=-\frac{1}{2}R_{nn}.\nn\\
 \label{Phi}
\end{eqnarray}
We note that $\Psi_4$ could not be represented by the Ricci tensor
and thus, it remains unconstrained in any metric theory of gravity.
Another relation is
\begin{equation}
R_{nknm}=R_{nm},
\end{equation}
while  a relation for the Ricci scalar is given by
\begin{equation} \label{risr}
R=-2R_{nk}=-2R_{nknk}=12\Psi_2,
\end{equation}
which implies  that the non-propagation of the Ricci scalar ($R=0$)
indicates  $\Psi_2=0$.

The naive  $E(2)$ classification of polarization waves under Lorentz
transformation~\cite{Eardley:1974nw} is useful to find gravitational
wave polarizations found in nCS modified gravity.
It is  given by  \\
$\bullet$ Class II$_6$: $\Psi_2\not=0$; \\
$\bullet$ Class III$_5$: $\Psi_2=0,~\Psi_3\not=0$; \\
$\bullet$ Class N$_3$: $\Psi_2=\Psi_3=0,~~\Psi_4\not=0,~~\Phi_{22}\not=0$; \\
$\bullet$ Class N$_2$: $\Psi_2=\Psi_3=\Phi_{22}=0,~~\Psi_4\not=0$; \\
$\bullet$ Class O$_1$: $\Psi_2=\Psi_3=\Psi_4=0,~~\Phi_{22}\not=0$; \\
$\bullet$ Class O$_0$: $\Psi_2=\Psi_3=\Psi_4=\Phi_{22}=0$. \\
The Einstein gravity of $R$ (equivalently, its equation
$R_{\mu\nu}=0$ and $R=0$) is of class N$_2$, while the EW gravity of
$R+\gamma(R_{\mu\nu}^2-R^2/3)$ may be classified by III$_5$ because
of the non-propagation of Ricci scalar ($R=0$). A quadratic gravity
of $R+\alpha R^2+\gamma R_{\mu\nu}^2$ may be categorized by class
II$_6$. However, we note that the number of DOF is seven for the EW
gravity and eight for the quadratic gravity in the metric tensor
formalism because of the presence of massive ghost modes.

The Brans-Dicke theory belongs to class N$_3$ because the
Brans-Dicke scalar provides $\Phi_{22}$~\cite{Alves:2010ms}. We note
that $\Phi_{22}$ and $\Psi_2$ correspond to a perpendicular scalar
mode (breathing mode) and to a longitudinal scalar mode. They are
arising from either the massive Brans-Dicke theory or the  metric
$f(R)$-gravity, in addition to an unspecified $\Psi_4$. These
theories are of class II$_6$ even though they have four polarization
modes with $\Psi_3=0$.

 The
$pp$-wave approach in nCS  modified gravity leads   to class
N$_2$~\cite{Sopuerta:2009iy}. If a theory is of class II$_6$ or
III$_5$, the corresponding amplitudes cannot be identified with
massless particle fields like $\Psi_4$ and $\Phi_{22}$. Furthermore,
the nCS modified gravity is a constraint theory regardless  of any
choice $\theta$. All solutions must satisfy the Pontryagin
constraint which translates into a reality condition of the
quadratic invariant for  the Weyl spinor ${\cal I}$ on the $E(2)$
classification~\cite{Grumiller:2007rv}
\begin{equation}
\Im({\cal I})=3\Im[\Psi_2^2]=0\to \Psi_2=0.
\end{equation}
It implies that any spacetime of Petrov types III, N, and O
automatically satisfies the Pontryagin constraint, while spacetimes
of Petrov type II could violate it.  We propose that if the nCS
gravity is a massive gravity theory, it belongs to III$_5$ but the
number of DOF is between five and two.  Hence,  we have to determine
what is the $E(2)$ class of nCS  modified gravity in the Minkowski
background really.

\section{nCS modified gravity}

Let us first consider the nCS modified gravity in four dimensions
whose action is given by
\begin{eqnarray}\label{csaction}
S_{\rm CS}=\frac{1}{2\kappa^2}\int
d^4x\sqrt{-g}\Bigg[R+\frac{\theta}{4}{}^*RR\Bigg],
\end{eqnarray}
where $\kappa^2=8\pi G$, $\theta$ is a nondynamical field, and
${}^*RR={}^*R^{\eta~\mu\nu}_{~\xi}R^{\xi}_{~\eta\mu\nu}$ is the
Pontryagin density with
\begin{eqnarray}
{}^*R^{\eta~\mu\nu}_{~\xi}
=\frac{1}{2}\epsilon^{\mu\nu\rho\sigma}R^{\eta}_{~\xi\rho\sigma}.
\end{eqnarray}
Varying for $g_{\mu\nu}$ on the action (\ref{csaction}), we find the
Einstein equation
\begin{eqnarray} \label{eequa}
R_{\mu\nu}-\frac{1}{2}g_{\mu\nu}R+C_{\mu\nu}=0.
\end{eqnarray}
Here, the four-dimensional Cotton tensor $C_{\mu\nu}$ is written by
\begin{eqnarray}
C_{\mu\nu}=\nabla_{\gamma}~\theta~\epsilon^{\gamma\rho\sigma}_{~~~(\mu}
\nabla_{|\sigma|}R_{\nu)\rho}+\frac{1}{2}\nabla_{\gamma}\nabla_{\rho}
~\theta~\epsilon_{(\nu}^{~~\gamma\sigma\delta}R^{\rho}_{~~\mu)\sigma\delta},
\label{cottont}
\end{eqnarray}
where the Cotton tensor is a traceless and symmetric tensor. We
stress that even though some works
\cite{Cardoso:2009pk,Moon:2011fw,Molina:2010fb} have focussed on the
latter term in (\ref{cottont}) for technical reasons, the former
term with $\theta=x/\mu$ shows really a 4D extension of the TMG in
three dimensions (see Eq. (\ref{maineq})). Hence, we keep the former
term in a sense that the nCS gravity is considered as a parity-odd
theory of massive gravity in four dimensions.

It is important  to note that applying $\nabla^{\mu}$ to
(\ref{eequa}) leads to a term of $\nabla^{\mu}C_{\mu\nu}=0$ which
can be rewritten as
\begin{eqnarray}\label{cont}
-\frac{\partial_{\nu}\theta}{8}{}^*RR~=~0,
\end{eqnarray}
being the  Pontryagin~ constraint of ${}^*RR=0$  for non-constant
$\theta$\footnote{For a constant $\theta$, Eq. (\ref{cont}) is
automatically satisfied. However, in this case, the Einstein
equation  (\ref{eequa}) reduces to that of  general relativity with
$C_{\mu\nu}=0$.}. Note that this is an outcome of the Bianchi
identity. In this work, we shall consider the Minkowski background
metric and scalar ansatz:
\begin{eqnarray}
&&ds^2=\eta_{\mu\nu}dx^{\mu}dx^{\nu}=-dt^2+dx^2+dy^2+dz^2,\label{flatm}\\
&&\hspace*{6.7em}\bar{\theta}=\frac{1}{\mu}x \label{flatt},
\end{eqnarray}
where $\mu$ is a constant related to a mass parameter. The scalar
ansatz is necessary to obtain a massive graviton propagating on the
Minkowski background.

We would like to mention a few things about the particular choice
(\ref{flatt}) of $\bar{\theta}$. The first is that $1/\mu$ plays a
role of {\it constant} coupling, which will be shown in Eq.
(\ref{lruv}). The second  is that it is needed to describe the plane
GWs propagating in the $+z$ direction, implying that all quantities
depend on the null retarded time of $(t-z)$ only. The other choice
of $\bar{\theta}=y/\mu$ is also possible, but it does not make a
significant difference when one compares with (\ref{flatt}). The
canonical embedding of $\bar{\theta}=t/\mu$, even though  it is
useful  for describing a massless graviton with 2 DOF in the metric
tensor formalism~\cite{Jackiw:2003pm}, is not suitable for
developing all modes in the nCS gravity.  This is because there are
no time derivatives in the linearized Cotton tensor for the
canonical embedding.

Now we introduce the metric perturbation as
$g_{\mu\nu}=\eta_{\mu\nu}+h_{\mu\nu}$, then the linearized equation
(\ref{eequa}) can be cast into the form
\begin{eqnarray}\label{lruv}
R^{{\rm L}}_{\mu\nu}-\frac{1}{2}\eta_{\mu\nu}R^{{\rm L}}
+\frac{1}{\mu}\bar{\epsilon}^{~x\rho\sigma}_{~~~~(\mu}
\partial_{|\sigma|}R^{{\rm L}}_{\nu)\rho} =0,
\end{eqnarray}
where $R^{{\rm L}}_{\mu\nu}$ and $R^{{\rm L}}$ are the linearized
Ricci tensor\footnote{We note that a quantity of $R^{{\rm
L}}_{\mu\nu}$ given in the Ricci tensor formalism is referred to a
boosted-up tensor of the massive graviton
\cite{Bergshoeff:2013vra,Bergshoeff:2012yz}, because it can be
written as $R^{{\rm L}}_{\mu\nu}=-\frac{1}{2}\square h^{\rm
TT}_{\mu\nu}$ under the transverse-traceless gauge.} and scalar
which consist of the components of the first-order metric
perturbation $h_{\mu\nu}$. The Levi-Civita symbol
$\bar{\epsilon}^{~\mu\nu\rho\sigma}$ is defined by
$\bar{\epsilon}^{txyz}=1$. On the other hand, taking the trace of
Eq.(\ref{lruv}) leads to
\begin{eqnarray}\label{r0}
R^{{\rm L}}=0,
\end{eqnarray}
which implies the non-propagation of Ricci scalar. Imposing $R^L=0$,
the linearized equation (\ref{lruv}) leads to the first-order time
equation for the Ricci tensor
\begin{eqnarray}\label{maineq}
R^{{\rm
L}}_{\mu\nu}+\frac{1}{\mu}\bar{\epsilon}^{~x\rho\sigma}_{~~~~(\mu}
\partial_{|\sigma|}R^{{\rm L}}_{\nu)\rho} \equiv R^{{\rm
L}}_{\mu\nu}+C^{\rm L}_{\mu\nu} =0,
\end{eqnarray}
which is our main equation for the linearized nCS modified gravity
around the Minkowski background. As far as we know, this is a newly
derived equation which is similar to the linearized equation given
in cosmological TMG~\cite{LSS}. It is easy to check that the
linearized Bianchi identity is satisfied as
\begin{equation} \label{bianch}
\partial^\mu R^{\rm L}_{\mu\nu}=0
\end{equation}
because of $\partial^\mu C^{\rm L}_{\mu\nu}=0$. Also, its traceless
equation is trivial: $R^{\rm L}=C^{\rm L}=0$.

 On the other hand, for the canonical embedding of
 $\bar{\theta}=t/\mu$, its linearized equation takes the form
\begin{eqnarray}\label{caemeq}
R^{{\rm
L}}_{\mu\nu}+\frac{1}{\mu}\bar{\epsilon}^{~t\rho\sigma}_{~~~~(\mu}
\partial_{|\sigma|}R^{{\rm L}}_{\nu)\rho}=0,
\end{eqnarray}
which is a genuine second-order time equation for the metric tensor
$h_{\mu\nu}$ (a first-order space equation for the Ricci tensor
$R^{{\rm L}}_{\mu\nu}$) because of $\sigma \not= t$.

\section{Polarization modes of GWs }

As was introduced in Sec.2, the NP formalism determines the number
of polarization states of GWs in four-dimensional gravity models.
Now we wish to find the number of GWs modes $\theta=x/\mu$ in nCS
modified gravity by using the NP formalism. To this end, we first
recall the null complex tetrad $\{{\bf k}, {\bf n},{\bf m},\bar{\bf
m}\}$ (\ref{null-t}) and the GWs set
$\{{\Psi_{2},\Psi_{3},\Psi_{4},\Phi_{22}}\}$ (\ref{Phi}). We point
out  that the vanishing of the linearized Ricci scalar (\ref{r0})
implies $\Psi_2=0$ which can be obtained by the relation
(\ref{risr}). It is worth noting that the condition of $\Psi_2=0$
also satisfies  the Pontryagin constraint ${}^*RR=0$, which
translates into a reality condition on a quadratic invariant of the
Weyl spinor  ${\cal I}$~\cite{Grumiller:2007rv}
\begin{eqnarray}
\Im({\cal I})&=&\Im(\Psi_0\Psi_4+3\Psi_2^2-3\Psi_1\Psi_3)\nonumber\\
&=&\Im(3\Psi_2^2)~=~0.\label{ii}
\end{eqnarray}
Here, we used $\Psi_0=\Psi_1=0$ because
 the two  Newman-Penrose scalars $\Psi_0$ and $\Psi_1$
vanish for  plane waves propagating along $+z$
axis~\cite{Eardley:1974nw}. On the other hand, it is noted that  the
GW mode $\Psi_{4}$ remains unconstrained in the nCS modified gravity
because it depends on the Riemann tensor $R_{n\tilde{m}n\tilde{m}}$.
Actually, it describes two massless GWs in the metric tensor
formalism.

We are now in a position to check whether the remaining modes
$\Psi_{3}$ and
 $\Phi_{22}$ (more explicitly, $R^{{\rm L}}_{nm}$, $R^{{\rm L}}_{n\tilde{m}}$,
$R^{{\rm L}}_{nn}$) are truly independent components. These are
modes of helicity $\pm1$ and 0. To this end, we realize that the
components of the linearized Ricci tensor in Eq. (\ref{maineq}) are
coupled to other components due to the Levi-Civita tensor.   It
turns out that $(t,t),~ (t,z),$ and $(z,z)$ components of Eq.
(\ref{maineq}) yield one relation
\begin{eqnarray}
&&R^{{\rm L}}_{tt}+2R^{{\rm L}}_{tz}+R^{{\rm
L}}_{zz}=-(\partial_t+\partial_z)\Big\{\frac{1}{\mu}(R^{{\rm
L}}_{yt}+R^{{\rm L}}_{yz})\Big\}\label{rel0},
\end{eqnarray}
which implies that the l.h.s of Eq. (\ref{rel0}) vanishes, because
acting an operation ($\partial_t+\partial_z \propto \partial_v$)  on
the linearized Ricci tensor with functions  of $(t-z \propto u)$
leads to zero.  This shows clearly a non-propagation of $\Phi_{22}$
($\Phi_{22}=0,R^{{\rm L}}_{nn}=0$) by observing the relations
\begin{eqnarray}
\Phi_{22}=-\frac{1}{2}R^{\rm L}_{nn},~~~R^{{\rm L}}_{nn}=R^{{\rm
L}}_{tt}+2R^{{\rm L}}_{tz}+R^{{\rm L}}_{zz}.
\end{eqnarray}
 After some manipulations, all remaining  components of Eq.
(\ref{maineq}) together with the  linearized  Bianchi identity
(\ref{bianch}) provide the other relation
\begin{eqnarray}
&&\Big\{1+\frac{1}{4\mu^2}(\partial_t^2-\partial_z^2)\Big\}\partial_t(R^{{\rm
L}}_{xt}+R^{{\rm L}}_{xz})=0,\label{relf}
\end{eqnarray}
which includes a case of the non-propagation of $R^{{\rm
L}}_{xt}+R^{{\rm L}}_{xz}$ (that is, $R^{{\rm L}}_{xt}+R^{{\rm
L}}_{xz}=0$) at the solution level. At this stage, introducing
$\Psi_3={\rm Re}[\Psi_3]+i{\rm Im}[\Psi_3]$ expressed in terms of
linearized Ricci tensor
\begin{eqnarray}
&&{\rm Re}[\Psi_{3}]=R^{{\rm L}}_{nm}+ R^{{\rm
L}}_{n\tilde{m}}=-(R^{{\rm L}}_{xt}+R^{{\rm L}}_{xz}),\label{rel1}\\
&&{\rm Im}[\Psi_{3}]=-i(R^{{\rm L}}_{nm}- R^{{\rm
L}}_{n\tilde{m}})=R^{{\rm L}}_{yt}+R^{{\rm L}}_{yz},\label{rel2}
\end{eqnarray}
(\ref{rel0}) and (\ref{relf}) indicate clearly that we have only one
independent component ${\rm Im}[\Psi_{3}]$. This is reminiscent of
the fact that  the TMG shows one DOF of a given helicity since the
gCS term is odd under parity, while the NMG has two DOF because it
belongs to a parity-even theory.

It seems that in the $\theta=x/\mu$ nCS modified gravity, there
exist three independent modes of GWs:
\begin{eqnarray}
&&\Big(\Psi_2=0,~~~\Phi_{22}=0,~~~{\rm Re}[\Psi_{3}]=0\Big)\nonumber\\
&&{\rm Re}[\Psi_{4}]\neq0,~~~{\rm Im}[\Psi_{4}]\neq0,~~~{\rm
Im}[\Psi_{3}]\neq0,
\end{eqnarray}
which correspond to the polarization modes $(a),~(b),$ and $(f)$ in
Fig. 1, respectively. Thus,  the nCS modified gravity might belong
to class III$_5$ with three DOF because $\Psi_2=0$ and
$\Psi_3\not=0$. On the other hand,  the other choice of
$\bar{\theta}=y/\mu$ might lead to the class III$_5$ with three DOF
since $\Psi_2=0$ and $\Psi_3\not=0~({\rm Re}[\Psi_{3}]\not=0,{\rm
Im}[\Psi_{3}]=0$) which indicates polarization modes $(a),~(b),$ and
$(e)$ in Fig. 1.

However, we would like to mention a weak point of the NP formalism
to represent  full modes  propagating on the Minkowski background in
the EW gravity. In the NP formalism, there is no constraint on
$\Psi_4$ in any metric theory of gravity because it is determined by
Riemann tensor $\Psi_4=-R_{n\tilde{m}n\tilde{m}}$ even for $R^{\rm
L}_{\mu\nu}=0$ of Einstein gravity.  Accordingly, two of seven GWs
in the EW gravity are given by $\tilde{\Psi}_4$, which describes a
massive tensor mode. The other five are given by $\Psi_4$, $\Psi_3$
and $\Phi_{22}$ in the NP formalism. On the other hand, in the Ricci
tensor formalism (see Sec.5 for details), the linearized Ricci
tensor equation (\ref{ein-w0}) indicates  that there exist five
($10-1-4=5$) DOF in the EW gravity, which consist of massive GWs
only. Here, 5 constraints are obtained  by requiring both the
traceless condition $R^{\rm L}=0$ and the transverse condition of
$\partial^\mu R^{\rm L}_{\mu\nu}=0$. Therefore, in the next section,
we use the Ricci tensor formalism to count the number of DOF of
propagating massive graviton in the $\theta=x/\mu$ nCS modified
gravity.

\section{Massive gravity equation}

In this section, we explore an explicit wave equation by employing
the Ricci tensor formalism. Thereby, we count the number of massive
modes in the nCS modified gravity.

 Before going further, we comment briefly on TMG and NMG.
 Let us first introduce a operator ${\cal
D}^{\mu}_{\nu}(m)=\delta^{\mu}_{\nu}
+\frac{1}{m}\epsilon_{\nu}^{~\alpha\mu}\partial_{\alpha}$ \cite{bht}
in three dimensions. Using the operator ${\cal D}$, the linearized
Einstein equation of the TMG takes the form $[{\cal D}(m)\delta
R]_{\mu\nu}=0$ with $\delta R=0$. Its second-order equation is given
by
\begin{equation} \label{3d-sec}
\Big[{\cal D}(-m){\cal D}(m) \delta R\Big]_{\mu\nu}=0 \to
(\Box-m^2)\delta R_{\mu\nu}=0,
\end{equation}
which is the linearized equation obtained from the NMG with $\delta
R=0$ and $\partial^\mu \delta R_{\mu\nu}=0$~\cite{bht}. This
indicates that the second-order equation (\ref{3d-sec}) with a
single mass-squared $m^2$  could be obtained from the first-order
equation naturally for the TMG. More schematically, we have
\begin{eqnarray}
\Big[D(-m_-)D(m_+)\delta R\Big]_{\mu\nu}=\left\{\begin{array}{ll}
m_+=m_-=m \hspace*{2.8em}\Rightarrow~\Box \delta R_{\mu\nu}-m^2
\delta R_{\mu\nu}=0\\
\\
m_+=m,~m_-\to\infty\\
~~~~~~~~{\rm or}\hspace*{5.91em}\Rightarrow~\delta
R_{\mu\nu}+\frac{1}{m}\bar{\epsilon}^{\rho\sigma}_{~~\mu}
\partial_{\rho}\delta R_{\sigma\nu}=0\label{ntmg} \\
m_-=-m,~m_+\to\infty \end{array}\right.
\end{eqnarray}
In this sense, the parity-odd TMG with one DOF, which corresponds to
a propagating mode with a mass $m_+$ or $m_-$, is considered as a
``square-root" of the parity-even NMG with two DOF.

Now we are in a position to develop a second-order wave equation in
four dimensions, which can describe a massive graviton propagation
explicitly. For this purpose, we first express the first-order
equation (\ref{maineq}) arisen from the nCS modified gravity by
introducing $D$ operators:
\begin{eqnarray}
D^{\lambda\lambda^{\prime}}_{\mu\nu}(\mu)~\equiv~\delta^{\lambda}_{(\mu}
\delta^{\lambda^{\prime}}_{\nu)}
+\frac{1}{\mu}\delta^{\lambda}_{(\mu}\bar\epsilon^{~x\lambda^{\prime}\sigma}_{~~~~~\nu)}
\partial_{\sigma},
\end{eqnarray}
where $(A,B)$ denotes the symmetrization of $(A,B)=(AB+BA)/2$. Using
the $D(\mu)$ operator, Eq. (\ref{maineq}) can be written compactly
as
\begin{eqnarray} \label{fir-eq}
D^{\lambda\lambda^{\prime}}_{\mu\nu}(\mu)R^{\rm
L}_{\lambda\lambda^{\prime}}~\equiv~\Big[D(\mu)R^{\rm
L}\Big]_{\mu\nu}~=~0.
\end{eqnarray}
Acting $\Big[D(-\mu)D(\mu)\Big]^{\lambda\lambda^{\prime}}_{\mu\nu}$
on $R^{\rm L}_{\lambda\lambda^{\prime}}$, we obtain a complicated
second-order equation
\begin{eqnarray}
&&\Big[D(-\mu)D(\mu)R^{\rm L}\Big]_{\mu\nu}\nonumber\\
 &&\hspace*{-0.9em}=~R^{\rm
L}_{\mu\nu}-\frac{1}{\mu^2}\Big[\Box R^{\rm L}_{\mu\nu}
+\frac{1}{2}\eta_{\mu\nu}\Box R^{\rm
L}_{xx}-\frac{1}{2}\partial_{\mu}\partial_{\nu}R^{\rm
L}_{xx}-\frac{3}{2}\delta^{x}_{(\mu}\Box R^{\rm L}_{\nu)x}\Big]~=~0.
\label{dd0}
\end{eqnarray}
This is not obviously a tensor wave equation that describes a
massive graviton propagating in the Minkowski background
\begin{equation} \label{ein-w0}
\Box R^{\rm L}_{\mu\nu}-\mu^2 R^{\rm L}_{\mu\nu}=0,
\end{equation}
which was obtained from the EW gravity (\ref{ewaction}) together
with $R^{\rm L}=0$, and $\partial^\mu R^{\rm
L}_{\mu\nu}=0$~\cite{Alves:2009eg}. Eq. (\ref{ein-w0}) is a standard
wave equation which describes five DOF of a massive graviton
propagating on the Minkowski background in the Ricci tensor
formalism. However, it is important to note that the $(x,x)$
component of Eq. (\ref{dd0}) leads simply to  $R^{\rm L}_{xx}=0$,
which kills one DOF. Therefore, we have four DOF ($5-1=4$) totally.
It turns out that finally, the tensor wave equation (\ref{dd0}) can
be classified into two types:
\begin{eqnarray}
\Big[D(-\mu)D(\mu)R^{\rm L}\Big]_{\mu\nu}=\left\{\begin{array}{ll}
\Box R^{\rm L}_{\mu\nu}-\mu^2 R^{\rm L}_{\mu\nu}=0, ~~(R^{\rm
L}_{tt},R^{\rm L}_{yt},R^{\rm L}_{zt},R^{\rm L}_{yy}
,R^{\rm L}_{yz},R^{\rm L}_{zz})~~~[{\rm mode}~1]\\
\\
\Box R^{\rm L}_{\mu\nu}-4\mu^2 R^{\rm L}_{\mu\nu}=0,~~~~~~~~(R^{\rm
L}_{tx},R^{\rm L}_{xy},R^{\rm L}_{xz})~~~~~~~~~~~~[{\rm
mode}~2].\label{soo4}\end{array}\right. \label{dd}
\end{eqnarray}
We would like to mention a few things observed from Eqs.
(\ref{maineq}) and (\ref{soo4}). Firstly,  the components of
linearized Ricci tensor in mode 1 do not couple to those in mode 2
\footnote{For example, one can check from the equation
(\ref{maineq}) that $R^{\rm L}_{tt}$ (one component of mode~1 in
Eq.(\ref{soo4})) can be expressed in terms of components $R^{\rm
L}_{tz}$ and $R^{\rm L}_{ty}$:
\begin{eqnarray}
R^{\rm L}_{tt}=\frac{1}{\mu}(\partial_{y}R^{\rm
L}_{tz}-\partial_{z}R^{\rm L}_{ty}).\nonumber
\end{eqnarray}}.
Secondly, we have one independent component of the Ricci tensor
coming from mode 1 when one requires five constraints from
(\ref{maineq}) for $(R^{\rm L}_{tt},R^{\rm L}_{yt},R^{\rm
L}_{zt},R^{\rm L}_{yy} ,R^{\rm L}_{yz},R^{\rm L}_{zz})$. The mode 2
is ruled out because its mass-squared is not $\mu^2$ but $4\mu^2$.
Thirdly, ${\rm Im}[\Psi_3]=R^{\rm L}_{yt}+R^{\rm L}_{yz}$ belongs to
the mode 1, while  ${\rm Re}[\Psi_3]=-(R^{\rm L}_{xt}+R^{\rm
L}_{xz})$ belongs to the mode 2.

By analogy of the TMG and NMG~(\ref{ntmg}), we propose that the
first second-order equation in (\ref{dd}) is related to the
first-order equation (\ref{maineq}) as follows:
\begin{eqnarray}
\Big[D(-m_1)D(m_2)R^{\rm L}\Big]_{\mu\nu}=\left\{\begin{array}{ll}
m_1=m_2=\mu \hspace*{3.5em}\Rightarrow~\Box R^{\rm L}_{\mu\nu}-\mu^2
R^{\rm
L}_{\mu\nu}=0\\
\\
m_1=-\mu,~m_2\to\infty \\
~~~~~~~~{\rm or}\hspace*{5.9em}\Rightarrow~R^{{\rm
L}}_{\mu\nu}+\frac{1}{\mu}\bar{\epsilon}^{~x\rho\sigma}_{~~~~(\mu}
\partial_{|\sigma|}R^{{\rm L}}_{\nu)\rho}=0, \label{ddd1}\\
m_2=\mu,~m_1\to\infty \end{array}\right.
\end{eqnarray}
which implies that when the $\theta=x/\mu$ nCS modified gravity is
considered as the parity-odd theory of massive gravity in four
dimensions, we have one  DOF which corresponds to a massive
propagating mode with mass $\mu$.

\section{Discussions}

In this work, we have considered the nCS modified gravity with
$\theta=x/\mu$ as a parity-odd model of massive gravity in four
dimensions like as the TMG~\cite{DJT2} in three dimensions. This
means that in this case, the nCS gravity is not a modified gravity
which mimics Einstein gravity of describing   two massless modes
$\Psi_4$. Here we have found one additional polarization mode of
${\rm Im}[\Psi_3$] by using the NP formalism. The presence of ${\rm
Im}[\Psi_3]$ implies that the nCS gravity provides three DOF (two
for massless mode and one for massive mode). This was compared to
the nCS gravity with the canonical embedding $\theta=t/\mu$ where
two massless modes are found only.

However, we have to point out an insufficiency of the NP formalism
to represent  full modes  in the EW gravity.  In the NP formalism,
there is no constraint on $\Psi_4$ in any metric theory of gravity
because it must be determined by Riemann tensor
$\Psi_4=-R_{n\tilde{m}n\tilde{m}}$, which corresponds to the
massless mode. In the case of the EW gravity, we have two modes of
$\tilde{\Psi}_4$ the massive ghost mode, in addition to five of
$\Psi_3$, $\Phi_{22}$, and $\Psi_4$~\cite{Alves:2009eg}. It shows
that the whole propagating modes are seven in the EW gravity.

As were previously shown in three-dimensional massive gravity
models~\cite{LSS,bht}, the third-and fourth-order gravity theories
become a first- and second-order gravity theories when one
introduces either the linearized Ricci tensor or Einstein tensor
(cosmological TMG and NMG) instead of metric tensor. A usage of
Ricci and Einstein tensors may be  a known way to avoid ghost
problem in higher-derivative gravity theories. Similarly, we have
introduced the Ricci tensor formalism to extract a massive mode
because we have regarded the $\theta=x/\mu$ nCS gravity as a massive
gravity. As a byproduct, we do not worry about the appearance of the
massive ghost states.

If the Ricci tensor was employed to describe the linearized Einstein
equation, the latter became the first-order equation (\ref{maineq})
which could be used to kill non-propagating modes in the nCS
gravity. In order to count the number of independent Ricci tensor
modes, we have to find the second-order wave equation (\ref{ein-w0})
together with the transverse-traceless condition found in the EW
gravity. As a result, we have found the tensor wave equation
(\ref{soo4}) for one independent mode which is compared to five from
the EW gravity~\cite{Alves:2009eg}.  Taking into account full modes
in the metric tensor formalism, the nCS gravity provides three,
while the EW gives us seven. In this sense, we may regard the nCS
gravity (parity-odd) as a square-root of the EW gravity
(parity-even), which is a four-dimensional version of ``TMG as a
square-root of NMG". In summary, comparing with the metric
formalism, both the NP and Ricci tensor formalisms are not suitable
for counting number of seven DOF in the EW gravity, while the Ricci
tensor formalism is not appropriate for finding three DOF in the
$\theta=x/\mu$ nCS gravity:

\vspace{0.5cm}
\begin{tabular}{|c|c|c|c|}
  \hline
  gravity models$\backslash$formalisms & metric tensor  & Ricci tensor  & NP \\
  \hline
  EW gravity & 7 DOF (2:massless+5:massive)& 5 (0+5) & 5 (2+3)\\ \hline
  $\theta=x/\mu$ nCS gravity & 3 (2+1) & 1 (0+1)& 3 (2+1)\\
  \hline
\end{tabular}
\vspace{0.5cm}

\vspace{10mm}

 {\bf Acknowledgments}

This work was supported by the National Research Foundation of Korea
(NRF) grant funded by the Korea government (MEST)
(No.2012-R1A1A2A10040499).

\end{document}